\documentclass[
	a4paper, 
	10pt, 
	twoside, 
]{LTJournalArticle}

\ExplSyntaxOn
\cs_gset:Npn \__first_footerline:
  { \group_begin: \small \sffamily \__short_authors: \group_end: }
\ExplSyntaxOff

\usepackage{todonotes}

\usepackage{graphicx} 
\usepackage{float}

\usepackage{natbib}
\usepackage[normalem]{ulem}
\usepackage[switch]{lineno}
\usepackage{subfigure}
\usepackage{graphicx}
\usepackage{caption}
\usepackage{subcaption}
\usepackage{float}
\usepackage{amsmath}
\usepackage[rightcaption]{sidecap}
\usepackage{todonotes}
\usepackage[table,xcdraw]{xcolor}
\usepackage{array}

\def\tsc#1{\csdef{#1}{\textsc{\lowercase{#1}}\xspace}}
\tsc{WGM}
\tsc{QE}
\tsc{EP}
\tsc{PMS}
\tsc{BEC}
\tsc{DE}

\begin{document}
\let\WriteBookmarks\relax
\def\floatpagepagefraction{1}
\def\textpagefraction{.001}

\newcommand{\rev}[1]{{\color{black}{#1}}}  
\newcommand{\alainold}[1]{{\color{black}{#1}}}
\newcommand{\alain}[1]{{\color{blue}{#1}}}

\runninghead{Destexhe, {\it CR Acad. Sci.}, 2026} 

\footertext{\today} 

\setcounter{page}{1} 


\title{A class of mean-field models to bridge molecular to brain scales\\ {\Large Application to model how anesthetics modify brain activity}}

\author{%
	Alain Destexhe\thanks{Correspondence: \href{mailto:alain.destexhe@cnrs.fr}{alain.destexhe@cnrs.fr} 
}
}

\date{\footnotesize
Paris-Saclay University, CNRS, Paris-Saclay Institute of Neuroscience (NeuroPSI), 91400 Saclay, France \\
\ \\ Submitted to {\it C.R. Acad. Sci}, \today
}

\renewcommand{\maketitlehookd}{%
    \begin{abstract}
        
        \noindent Predicting how molecular changes affect large-scale brain activity is a difficult task because of the lack of appropriate methods to link scales.  In this perspective, we review a class of mean-field models that can integrate biophysical details such as synaptic receptors or membrane ion channels.  This leads to a multi-scale modeling approach that can be used to evaluate how microscopic changes can impact macroscopic brain activity.  This approach is illustrated here for the case of anesthesia, where changes at the level of specific synaptic receptors can lead to a global change in brain activity and a disconnection from external inputs.  This is only possible using mean-field models that can include enough detail about the microscopic biophysical properties.  This biophysically-based mean-field approach could be generalized to study cellular or molecular origins of brain diseases, or to better understand how drugs acting at microscopic scales can influence global brain activity.  Biophysical mean-field models also link different fields of neuroscience, from molecular studies to brain imaging.
        
	\end{abstract}
}




\maketitle

\section{Introduction}


Numerous diseases and pharmacological compounds act at the molecular level, inducing changes in brain states and large-scale brain activity. However, there is currently no framework capable of assessing how molecular alterations—such as changes in synaptic receptor function—translate into large-scale brain dynamics, despite the importance of this question for understanding drug mechanisms and endogenous regulatory processes such as neuromodulation. One major challenge is that molecular interactions within neurons and the emergence of large-scale brain activity occur at vastly different spatial and temporal scales. The human brain contains approximately 85 billion neurons connected by an estimated $10^{15}$ synapses, making the construction of a whole-brain model with single-neuron resolution computationally infeasible. Bridging these disparate scales therefore remains a fundamental challenge in computational neuroscience. In this perspective paper, we review a class of mean-field models that provide a framework for linking molecular processes to whole-brain dynamics, and argue that they provide an important bridge, not only between scales but also between different disciplines of neuroscience.

\section{Biophysical mean-field models} \label{sec-newclass}

Linking scales is only possible if there is a formalism that can integrate microscopic information to evaluate how it can influence dynamics at larger scale.  A typical approach for this bottom-up construction is to design mean-field models describing the mesoscale dynamics of neural populations.  However, classic mean-field approaches are analytic or linear, and thus are confined to relatively simple models, that cannot integrate biophysical mechanisms and their nonlinearities.  Yet, such nonlinearities are of primary importance for the emerging properties at large scale, and should not be neglected.  We review here a class of mean-field models capable of integrating nonlinear biophysical mechanisms.

\subsection{Master Equation-based mean-field model}

A mean-field approach introduced a few years ago \citep{boustani2009master,zerlaut2018modeling} can integrate biophysical properties and thus complex models and nonlinear interactions.  This mean-field formalism was derived from a Master Equation \citep{boustani2009master}, and can be written as :
\begin{eqnarray} \label{mean-field_eq}
T\frac{\partial \nu_{\mu}}{\partial t}& = & (F_{\mu}-\nu_{\mu})+\frac{1}{2}c_{\lambda\eta}\frac{\partial^2 F_{\mu}}{\partial\nu_{\lambda}\partial\nu_{\eta}}   \\
T\frac{\partial c_{\lambda\eta}}{\partial t} & = & \delta_{\lambda\eta}\frac{F_{\lambda}(1/T-F_{\eta})}{N_{\lambda}}+(F_{\lambda}-\nu_{\lambda})(F_{\eta}-\nu_{\eta})
\nonumber \\
& & +\frac{\partial F_{\lambda}}{\partial \nu_{\mu}}c_{\eta\mu}+\frac{\partial F_{\eta}}{\partial \nu_{\mu}}c_{\lambda\mu}-2c_{\lambda \eta} \\
\frac{\partial W}{\partial t} & = & \frac{1}{\tau_w}(a (\mu_V(\nu_e, \nu_i, W) - E_L) - W) + b \nu_e,
\label{popadapt} 
 \end{eqnarray}
\noindent
where $\mu=\{e, i\}$ is the neural population index (excitatory or inhibitory), $ \nu_{\mu}$ the mean firing rate of the corresponding population, $c_{\lambda\eta} $ is the covariance between populations $\lambda$ and $\eta$,  $W$ is the population-level adaptation, $T$ is the mean-field characteristic time constant and $F_{\mu}$ is the transfer function of neuron type ${\mu}$ (see Section~\ref{sec-transf}).  For a detailed derivation of these equations and a discussion of
the associated closure assumptions and approximations, see \cite{Bossard2026}.

The first-order version of the mean-field is obtained by disregarding the covariance term dynamics in Equation~\ref{mean-field_eq} and can be written as:
\begin{eqnarray}  \label{singlenodeMF}
T \frac{d \nu_e}{dt} & = &  
\mathcal{F}_e(\nu_e, \nu_i, W ) -  \nu_e  \\
T \frac{d \nu_i}{dt} & = &  \mathcal{F}_i(\nu_e, \nu_i , W) -  \nu_i \\
\frac{\partial W}{\partial t} & = & \frac{1}{\tau_w}(a (\mu_V(\nu_e, \nu_i, W) - E_L) - W) + b \nu_e
\end{eqnarray}

\subsection{Semi-analytic approach} \label{sec-transf}

The function $F_{\mu}=F_{\mu}(\nu_e,\nu_i,W)$ is the transfer function (TF) of neuron type $\mu$, meaning its output firing rate for all combinations of inhibitory and excitatory inputs with rates $\nu_e$ and $\nu_i$, and adaptation level $W$. The TF is central to define the mean-field, and is in general not known analytically.  However, a definite advancement was to derive a semi-analytic approach \citep{zerlautHeterogeneousFiringRate2016a,zerlaut2018modeling}, where the output firing rate of a neuron can be written as a function of its mean subthreshold membrane voltage $\mu_V$, its standard deviation $\sigma_V$, and its correlation decay time $\tau_V$: 
\begin{equation} \label{eq:tf_fout}
F = \nu_{\text{out}} = \frac{1}{2 \tau_V} \cdot \text{Erfc} \left( \frac{V_{\text{thr}}^{\text{eff}} - \mu_V}{\sqrt{2} \sigma_V} \right) ,
\end{equation}
where ($\mu_V,\sigma_V, \tau_V $) are calculated as a function of the input
firing rates ($\nu_E, \nu_I$ ) and the adaptation intensity $W$ following the equations described in \cite{divolo2019biologically}. {Specifically, the mean membrane potential is calculated as the stationary solution under static conductances driven by the average synaptic input generated by firing rates ($\nu_E, \nu_I$ ). This input determines the mean ($\mu_{G_e}, \mu_{G_i}$) and standard deviation ($\sigma_{G_e},\sigma_{G_i}$) of conductances for excitatory and inhibitory processes (see details in Appendix~1).   

The basis of the semi-analytic approach is that if the firing threshold can be defined appropriately (see Appendix~1), then the analytic template (Eq.~\ref{eq:tf_fout}) can fit the firing response of many complex neuron types.  This is obtained by fitting the template with a varying threshold to numerical simulations of single-neuron dynamics.  The mean-field model obtained thus remains analytical, only its parameters are fit numerically.   We discuss below the properties of the mean-field models obtained using this semi-analytic framework.

\section{Properties of the biophysical mean-field}

In this section, we discuss the properties of the Master-Equation based mean-field model and its semi-analytic formulation, which we will refer as {\it biophysical mean-field}.

\subsection{Applicability to a wide range of models}

The biophysical mean-field formalism as presented in Section~\ref{sec-newclass} was found to apply to various types of neurons.  It was first applied to networks made of two cell types, regular-spiking (RS) excitatory cells and fast-spiking (FS) inhibitory cells \citep{zerlaut2018modeling}, two of the main cell classes found in the nervous system.  These were described by the adaptive exponential (AdEx) integrate and fire neuron \citep{brette2005adaptive}. Using the AdEx model, it was possible to model propagation phenomena in visual cortex \citep{zerlaut2018modeling} and different cortical network states, such as asynchronous-irregular dynamics and slow oscillations with Up and Down states \citep{divolo2019biologically}.  These states can be simulated by AdEx-based spiking network simulations and the biophysical mean-field accurately matches the network simulations for different network states (\cite{divolo2019biologically}; Fig.~\ref{fig:states}).

This approach, using AdEx models, was further explored for other brain regions.  The biophysical mean-field could be derived from thalamic neurons \citep{overwiening2024multi}, which was non-trivial because these neurons exhibit bursts of action potentials.  Using numerical simulations constrained by {\it in vitro} recordings of thalamic neurons \citep{wolfart2005}, it was possible to formulate the TF for thalamic relay cells under {\it in vivo} conditions, and define a semi-analytic mean-field specific for thalamic neurons \citep{overwiening2024multi}.  As for cortex, the mean-field was compared to spiking network simulations, both for spontaneous activity and evoked responses.

Other two-dimensional integrate and fire models were used similarly to the AdEx model, to define biophysical mean-field models of different brain regions, such as the striatum \citep{tesler2026multiscale}, hippocampus CA1 region \citep{Tesler2024Hippo} and cerebellum \citep{lorenzi2023multi}.  In each case, the firing properties of the neurons were captured by a nonlinear integrate and fire model, and the transfer function was fitted to numerical simulations of this model.  The mean-field models obtained were all validated against spiking network simulations, for spontaneous and evoked activity.

\begin{figure*}[ht!]  

\centering
\includegraphics[width=0.55\textwidth]{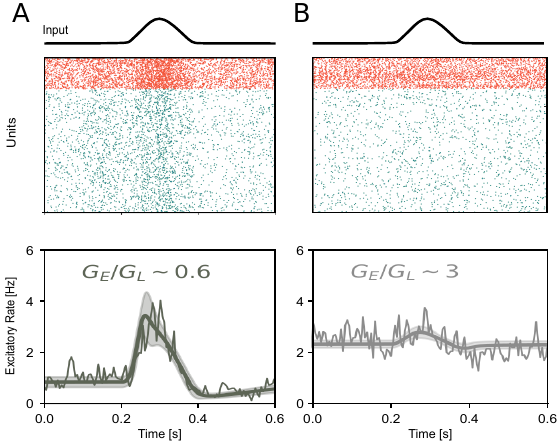}

\caption{Responsiveness a network of spiking (AdEx) neurons in two different asynchronous states. Top: rasters of spiking activity (red, FS cells; blue, RS cells).  Bottom: mean firing activity computed from the population of RS cells.  A Gaussian-shaped excitatory stimulus was given (top traces), and the response of the network was monitored.  A and B show two different combinations of parameters, a low-conductance state (A) and a high-conductance state (B).  The very different responsiveness was well captured by the mean-field model (continuous lines in bottom panels).  Modified from \cite{divolo2019biologically} where details can be obtained.}

\label{fig:states} 
\end{figure*}    

Finally, it must be mentioned that the biophysical mean-field approach can also be applied to more complex cases, such as the Hodgkin-Huxley (HH) model \citep{carluMeanfieldApproachDynamics2020} and even to living neurons \citep{zerlautHeterogeneousFiringRate2016a}.  In the case of the HH model, networks of spiking neurons were formed using HH models for RS and FS cells, and the network exhibited asynchronous-irregular activity states.  The biophysical mean-field was derived by computing and fitting the TF to HH model neurons.  The mean-field model obtained could capture the spontaneous activity and evoked responses of HH networks.  

In the case of living neurons, perforated patch-clamp experiments were performed to scan the firing properties of Layer~5 pyramidal neurons from mouse visual cortex \citep{zerlautHeterogeneousFiringRate2016a}.  This scanning was performed using dynamic-clamp injection of excitatory and inhibitory synaptic conductances.  Here again, the TF could be calculated and fit to the firing patterns recorded.  However, a considerable cell-to-cell variability was observed, where neurons with apparently similar Layer~5 morphology can have very different excitability properties.  This motivated the design of mean-field models of heterogenous networks of neurons (see details in \cite{di2021optimal}).

\subsection{Nonlinear interactions}

The availability of a formalism that can capture complex neural models makes it possible to include essential biophysical characteristics of neurons, which are typically absent in simple models.  One of these characteristics is the conductance-based nature of synaptic inputs.  The nonlinearity is here due to the multiplication between the conductance and the voltage (driving force) terms, and has many consequences on neurons.  For example, it is well known that shunting inhibition, a typical conductance effect, is essential to explain visual responses \citep{Borg-Graham98}.  It was also found that conductances profoundly change the integrative properties of neurons under {\it in vivo} conditions \cite{Destexhe2003}.   Moreover, as can be seen in Fig.~\ref{fig:states}, different network states with different level of synaptic conductances have a different response to external inputs, a situation which is impossible to capture with current-based models.

It is important to keep in mind that the most widely used mean-field models are based on a current-based approximation of synaptic conductances, where the synaptic currents have no driving force, and thus the summation of synaptic inputs is linear.  This is the case for the Montbrio model \citep{Montbrio2015}, which derives an exact mean-field for a network of quadratic integrate-and-fire neurons with current-based synapses.  The Wong-Wang model \citep{WongWang2006} is derived from the leaky integrate-and-fire model, and uses a form inspired from the well-known Wilson-Cowan model \citep{WilsonCowan72}.  In contrast, the mean-field models we discuss here can integrate biophysical aspects such as conductances, synaptic receptors, and multiple neuronal firing patterns, thus departing from linear input summation, and capturing the nonlinearities present in neurons to a larger extent.

\begin{figure*}[ht!]   

\centering
\includegraphics[width=1\textwidth]{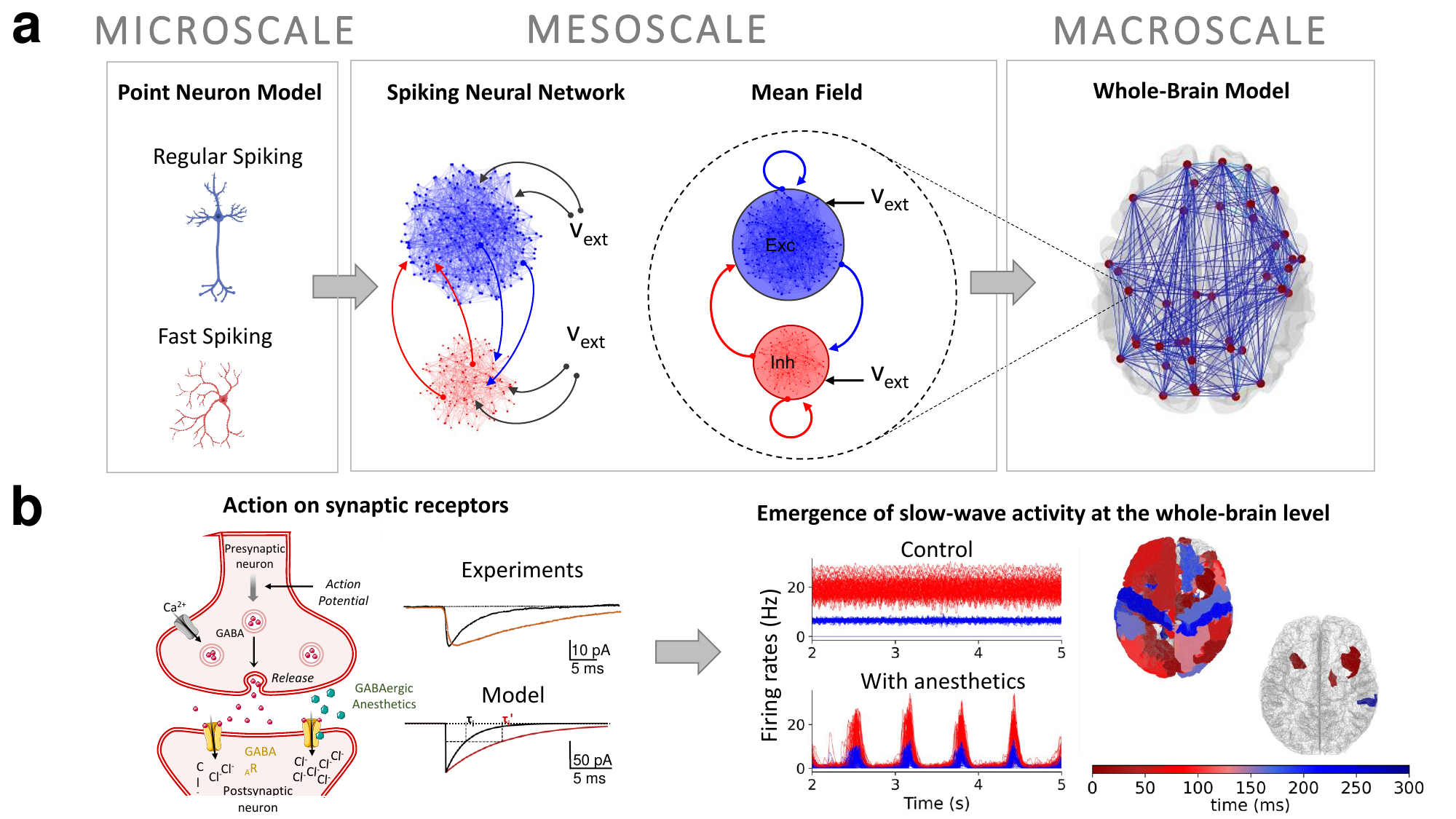}

\caption{Illustration of the framework to link microscopic and macroscopic scales of the brain.  A. Scheme of the different scales involved, from single neurons (Microscale), networks of neurons and populations (Mesoscale), up to the whole-brain (Macroscale).  B. Example of modeling the action of anesthetics (here on the GABA$_A$ receptor; left) which leads to the emergence of slow-wave activity (right panels).  The slow-wave state displays a diminshed evoked response complexity (rightmost panels).  Modified from \cite{sachaComputationalApproachEvaluate2025a} where all details can be found.}

\label{fig:architec}
\end{figure*}   



\subsection{Second-order and finite size}

A main originality of the mean-field described in Eqs.~\ref{mean-field_eq} is that it is second-order, comprising not only the mean firing rates but also the covariances as variables.  A second-order mean-field takes into account fluctuations dynamically, as well as time-varying correlations, while a first-order mean-field does not.  Correlations and fluctuations were shown to be of primary importance in conditions where there is a balance between excitation and inhibition \citep{vanVreeswijk-Sompolinsky1996}.  Indeed, the aim of the mean-field formalism was precisely to model balanced network states \citep{boustani2009master}, which motivated the use of a second-order mean-field.

Linear mean-field models \citep{Montbrio2015} summate synaptic inputs linearly and therefore probably fail to distinguish between all types of temporal structure of the firing activity.  For example, the shunting aspect of the conductance can have strong consequences \citep{Borg-Graham98,Destexhe2003} which are missed by a linear model.  The biophysical mean-field model, because of its second-order and conductance-based nature, does capture the temporal structure of the response of the population (Fig.~\ref{fig:states}).  We believe this is an essential property to meaningfully model the interactions between neuronal populations.  Note that this comparison of the mean-field response and the response of the spiking network to afferent synaptic stimulation is rarely done, despite its importance for modeling interacting populations.

Another originality of the second-order mean-field is that it was designed as a finite-size model.  The size of the network appears in the parameters of the second-order terms in Eqs.~\ref{mean-field_eq}.  This aspect also departs from classic mean-field models, which are only valid for very large network sizes.  For instance, the Montbrio model \citep{Montbrio2015} is formally exact only for a network of infinite size.  By contrast, the second-order mean-field model is valid for moderate network sizes, although this aspect was never studied explicitly (for a discussion, see \cite{boustani2009master}).

\section{Large-scale neural systems}

We now illustrate the applicability of the biophysical mean-field approach to build multiscale models, with the aim of relating microscopic aspects, such as ionic conductances or synaptic receptors, to the emergence of large-scale activity changes in the brain.  To do this, we use biophysical mean-field models capable of integrating details about the synaptic receptors, as well as electrophysiological data on the firing patterns and excitability of different neuron types.  These, in turn can be used for large scale simulations of the brain, as illustrated in Fig.~\ref{fig:architec}.  The extension of this local mean-field framework to
receptor-aware whole-brain simulations, together with its numerical implementation and computational cost, is discussed in detail in \cite{Bossard2026}.

This multiscale modeling approach was applied recently to model the effect of anesthetic agents on brain activity \citep{sachaComputationalApproachEvaluate2025a}.  We integrated the effect of anesthetics targeting inhibitory (GABA$_A$) receptors, such as propofol, isoflurane/sevoflurane or barbiturates, as well as anesthetics such as ketamine acting on excitatory (NMDA) synaptic receptors in cortical neurons.  Integrating these receptors in mean-field models leads to whole-brain models where one can study the variation of microscopic parameters corresponding to these receptors.  We found that modifications on these receptors can switch brain activity to generalized slow-wave patterns, as observed experimentally in deep anesthesia. To validate the models obtained, we used two experimentally observed properties of anesthetized states, first a reduced brain responsiveness to external stimuli \citep{massimini2005}, and second, a shift of the functional connectivity towards structural connectivity \citep{barttfeld2015signature}.  Both properties were present in the model, which indicates that its whole-brain dynamics share many similarities with anesthetized brain states.

The same approach was followed to simulate the action of psychedelic drugs such as psilocybin \citep{Martin2026}.  In this case, it was found that the main receptor type affected by psilocybin, the serotonin 5-HT$_{2A}$ receptor, acts through the depolarization of pyramidal cells and interneurons by reducing their leak K$^+$ conductance (reviewed in \cite{Andrade2011}).  This leak K$^+$ conductance was implemented in the model, for both RS and FS cells, and the same procedure as in \cite{sachaComputationalApproachEvaluate2025a} was followed to construct a whole-brain model.  The main effect of 5-HT$_{2A}$ receptor stimulation in the model was to increase the intensity and complexity of the fluctuations of activity, matching experimental observations \citep{Schartner2017,Murray2024}.

It is important to mention that the biophysical mean-field approach can also be used to infer possible mechanisms underlying experimental observations.  This was the case for propagating waves in monkey visual cortex, which were characterized using voltage-sensitive dye imaging \citep{Muller2014}.  Evoking a collision between two propagating waves revealed a suppressive component in their interaction \citep{Chemla2019}.  Modeling V1 using a network of mean-field models could reproduce these results, and the model predicted that the suppression was due to the difference of gain between excitatory and inhibitory neurons (see details in \cite{Chemla2019}).  This is not only a nice example of an emerging consequence of the fact of having two cell types, RS and FS cells, but it is also an example of the power of biophysical mean-field models: they can be reverse-engineered to identify possible biophysical causes of an observed macroscopic effect.

\section{Discussion}

In this paper, we presented an overview of a class of mean-field models capable of integrating biophysical details of neurons and synaptic interactions through multiple synaptic receptors.  We discuss here the consequences and possible impact of this type of model to build a new generation of truly multi-scale models of brain function and brain pathologies.

As we have reviewed in the paper, the semi-analytic approach has revolutionized the applicability of mean-field models, which were initially limited to simple integrate and fire models \citep{WongWang2006} or their quadratic versions \citep{Montbrio2015}.  These previous mean-field models used current-based (linear) approximations of synaptic interactions, which confines them to the linear integration of synaptic inputs, while in reality the integration is highly nonlinear (e.g, see discussion in \cite{Destexhe2003}).  Moreover, neurons are endowed with complex and nonlinear firing properties well beyond the integrate-and-fire paradigm, and overly simple mean-field models are not capable of integrating such complex firing properties, and thus cannot be used to evaluate their possible consequences at the large scale.

The initial Master Equation-based mean-field could capture the population dynamics of integrate and fire networks with conductance-based synapses \citep{boustani2009master}.  It was later extended using a semi-analytic approach \citep{zerlautHeterogeneousFiringRate2016a,zerlaut2018modeling}, which considerably expanded the applicability of the formalism to derive biophysical mean-field models of various neural systems with complex firing responses \citep{zerlaut2018modeling,divolo2019biologically,carluMeanfieldApproachDynamics2020,di2021optimal,lorenzi2023multi,overwiening2024multi,Tesler2024Hippo,tesler2026multiscale} and was also applied to capture the firing responses of pyramidal neurons in slices \citep{zerlautHeterogeneousFiringRate2016a}.  Such a wide range of applicability was not possible before, and while the tendency in theoretical neuroscience is often to consider simplified and tractable models, it is now possible to integrate this complexity while remaining tractable (since the biophysical mean-field remains analytic).  Importantly, it allows one to explicitly take into account the electrophysiological measurements in different neuron types in the nervous system.  Therefore, we believe this represents a significant advance for computational neuroscience, because one can build large-scale models that explicitly integrate experimental measurements of the excitability profile of different cell types. 

Where this advancement becomes particularly relevant is when we consider interactions over larger scales than the local networks.  Here, the biophysical mean-field approach has enabled us to directly predict how changes at the microscopic level can generate or alter brain activity at macroscopic scales.  This can be at the level of millimeters, with propagating waves in V1 \citep{Chemla2019}, or the brain-scale emergence of slow-wave activity arising from simulated interactions between anesthetics on synaptic receptors \citep{sachaComputationalApproachEvaluate2025a}.  In both cases, it was possible to relate how microscopic properties influence aspects of the macroscopic activity of brain tissue.

This approach offers perspectives for future studies.  First, the design of region-specific biophysical mean-field models should be pursued, with the aim to build whole-brain models where each region is described by its own specific model, respecting its particular firing and excitability properties.  This is true not only for regions such as basal ganglia, thalamus or cerebellum, but also for different brain areas within the cerebral cortex.  The availability of such a ``multi-region'' whole-brain model would be a very useful tool for computational studies.  Second, the same approach as \cite{sachaComputationalApproachEvaluate2025a} could be used to investigate other paradigms such as hallucinogenic drugs or brain pathologies arising from malfunction of synaptic receptors or other biophysical parameters.  Perhaps most importantly, it could be used to also investigate numerically how the pathological situation can be reversed, which could provide useful predictions for the treatment of brain pathologies.

Finally, let us mention that this approach is part of a long-term goal to use computational neuroscience tools to help investigate subjects difficult to study with traditional experimental neuroscience tools.  Known examples are how to predict the collective behavior of networks of neurons, given their complex nonlinear properties.  Similarly here, we constructed a framework to generate ``biologically realistic'' population-level models of brain activity, by integrating molecular information and biophysical properties, with an aim to predict consequences at the level of the whole brain.  This approach therefore not only links scales but it also links different fields of neuroscience, such as molecular neuroscience and electrophysiology with large-scale investigations such as brain imaging and connectomics.  

\subsection*{Acknowledgments}

Research supported by the CNRS, the ANR (FLAG-ERA BrainAct project; CR-CNS ImpactCom project) and the European Union (Virtual Brain Twin project 101137289, EBRAINS-2.0 project 101147319). I thank all the lab colleagues and collaborators for their great work and insightful discussions.

\appendix

\section*{Appendix 1: Derivation of the transfer function for conductance-based models}

Assuming Poissonian spike statistics (arising from asynchronous irregular dynamics), the mean and variance of synaptic conductances are expressed as:
\begin{align}
\mu_{G_s}(\nu_e, \nu_i) &= \nu_s K_s \tau_s Q_s, \label{eq:mu_g} \\
\sigma_{G_s}(\nu_e, \nu_i) &= Q_s \sqrt{\frac{\nu_s K_s \tau_s}{2}}. \label{eq:sd_g}
\end{align}
where $K_s = pN_s$ and $s = \{e, i\}$

The mean input conductance $(\mu_G)$ and the effective membrane time constant $(\tau_m^{\text{eff}})$ are given by:
\begin{align}
\mu_G(\nu_e, \nu_i) &= \mu_{G_e} + \mu_{G_i} + g_L, \\
\tau_m^{\text{eff}}(\nu_e, \nu_i) &= \frac{C_m}{\mu_G}.
\end{align}

The average membrane potential for a given adaptation current $w$ is:
\begin{eqnarray}
\mu_V(\nu_e, \nu_i, w) = \frac{\mu_{G_e}E_e + \mu_{G_i}E_i + g_L E_L - w}{\mu_G}.
\end{eqnarray}

The standard deviation $(\sigma_V)$ and time constant $(\tau_V)$ of voltage fluctuations are:
\begin{align}
\sigma_V(\nu_e, \nu_i) &= \sqrt{\sum_s K_s \nu_s \frac{(U_s \cdot \tau_s)^2}{2(\tau_m^{\text{eff}} + \tau_s)}}, \\
\tau_V(\nu_e, \nu_i) &= \frac{\sum_s \left(K_s \nu_s (U_s \cdot \tau_s)^2 \right)}{\sum_s \left(K_s \nu_s \frac{(U_s \cdot \tau_s)^2}{\tau_m^{\text{eff}} + \tau_s}\right)}.
\end{align}
where  $s = \{e, i\}$ and $U_s = \frac{Q_s,\mu_G} (E_s - \mu_V)$.

}
The $V_{\text{thr}}^{\text{eff}}$ in Eq~\ref{eq:tf_fout} is the phenomenological spike threshold, which depends on voltage mean, standard deviation and time constant, according to the second-order polynomial: 
\begin{equation} \label{eq:veff}
\begin{aligned}
V_{\text{eff}}^{\text{thr}}(\mu_V, \sigma_V, \tau_V^N) = P_0 + \sum_{x \in \{\mu_V, \sigma_V, \tau_V^N\}} P_x \left(\frac{x - x_0}{\delta x_0}\right) \\
 + \sum_{x, y \in \{\mu_V, \sigma_V, \tau_V^N\}^2} P_{xy} \left(\frac{x - x_0}{\delta x_0}\right) \left(\frac{y - y_0}{\delta y_0}\right),
\end{aligned}
\end{equation}
where  \(\tau_V^N = \frac{\tau_V g_l}{C_m}\) a dimensionless quantity. The polynomial coefficients of $P$ are determined through a fitting of the TF template to the output firing rate of individual neuron simulations, varying both inhibitory and excitatory inputs. Examples values for the normalization of the fluctuation regime were set following previous work \citep{zerlaut2018modeling, divolo2019biologically}:  $\mu_V^0 =$ -60~mV,  $\sigma_V^0 =$ 4~mV,  $(\tau_V^N)^0 =$ 0.5,  $\delta \mu_V^0 =$ 10~mV,  $\delta \sigma_V^0 =$ 6~mV and $\delta (\tau_V^N)^0 =$ 1. The fit is performed for each of the cell types included in the network.


\bibliographystyle{bib-names}

\bibliography{refsemerg}

\end{document}